\documentclass[preprintnumbers,superscriptaddress,amsmath,amssymb,nofootinbib,twocolumn,showpacs]{revtex4-1}

\usepackage[dvips]{graphicx}
\usepackage{amsmath,amssymb,mathrsfs}
\usepackage{bm}
\usepackage{times}
\usepackage{epsfig}
\usepackage{verbatim}
\usepackage{bm}
\usepackage[utf8]{inputenc}
\usepackage{graphics}
\usepackage{graphicx,epsfig,amssymb,amsmath,color, cancel}
\usepackage{soul}
\usepackage[normalem]{ulem}
\usepackage[breaklinks=true, colorlinks=true, linkcolor=magenta, urlcolor=blue, citecolor=blue]{{hyperref}}

\definecolor{vert1}{cmyk}{0.92,0,0.59,0.25}
\definecolor{bleu1}{cmyk}{1,1,0,0.3}
\definecolor{bleu2}{cmyk}{1,1,0,0.6}
\definecolor{bleu3}{cmyk}{1,1,0,0.1}
\definecolor{magenta}{cmyk}{0.,1.,0.,0.1}

\usepackage[english]{babel}
\usepackage{amsfonts,amsmath,amssymb,amsthm,mathtools}
\usepackage{comment,enumerate,footnote,graphicx,subfloat,relsize}
\usepackage{array,tabularx,tabu,multirow,framed,afterpage}
\usepackage[usenames,dvipsnames]{xcolor}
\usepackage{cleveref}
\crefformat{pluralequation}{#2\black{eqs.~(}#1\black{)}#3}
\Crefformat{pluralequation}{#2\black{Equations~(}#1\black{)}#3}
\crefformat{pluralfigure}{#2\black{figs.~}#1#3}
\Crefformat{pluralfigure}{#2\black{Figures~}#1#3}

\graphicspath{{./}{Figs/}}

\newcommand{\beq}{\begin{equation}}
\newcommand{\eeq}{\end{equation}}
\newcommand{\ben}{\begin{eqnarray}}
\newcommand{\een}{\end{eqnarray}}
\newcommand{\bi}{\begin{itemize}}
\newcommand{\ei}{\end{itemize}}

\newcommand{\vecx}{\mbox{$\vec{x}$}}

\newcommand{\fsol}{{\ifmmode f_{\odot} \else $f_{\odot}$\fi}}

\newcommand{\sqgev}{{\ifmmode {\rm GeV}^2 \else ${\rm GeV}^2$\fi}}

\newcommand{\sr}{{\ifmmode {\rm sr} \else ${\rm sr}$\fi}}
\newcommand{\invsr}{{\ifmmode {\rm sr}^{-1} \else ${\rm sr}^{-1}$\fi}}

\newcommand{\scnd}{{\ifmmode {\rm s} \else ${\rm s}$\fi}}
\newcommand{\invscnd}{{\ifmmode {\rm s}^{-1} \else ${\rm s}^{-1}$\fi}}

\newcommand{\kpc}{{\ifmmode {\rm kpc} \else ${\rm kpc}$\fi}}
\newcommand{\invkpc}{{\ifmmode {\rm kpc}^{-1} \else ${\rm kpc}^{-1}$\fi}}

\newcommand{\sqkpc}{{\ifmmode {\rm kpc}^{2} \else ${\rm kpc}^{2}$\fi}}
\newcommand{\invsqkpc}{{\ifmmode {\rm kpc}^{-2} \else ${\rm kpc}^{-2}$\fi}}

\newcommand{\cm}{{\ifmmode {\rm cm} \else ${\rm cm}$\fi}}
\newcommand{\invcm}{{\ifmmode {\rm cm}^{-1} \else ${\rm cm}^{-1}$\fi}}

\newcommand{\sqcm}{{\ifmmode {\rm cm}^2 \else ${\rm cm}^2$\fi}}
\newcommand{\invsqcm}{{\ifmmode {\rm cm}^{-2} \else ${\rm cm}^{-2}$\fi}}

\newcommand{\meter}{{\ifmmode {\rm m} \else ${\rm m}$\fi}}
\newcommand{\invmeter}{{\ifmmode {\rm m}^{-1} \else ${\rm m}^{-1}$\fi}}

\newcommand{\sqmeter}{{\ifmmode {\rm m}^2 \else ${\rm m}^2$\fi}}
\newcommand{\invsqmeter}{{\ifmmode {\rm m}^{-2} \else ${\rm m}^{-2}$\fi}}

\newcommand{\lcdm}{{\ifmmode \Lambda{\rm CDM} \else $\Lambda{\rm CDM}$\fi}}

\newcommand{\Rvirh}{{\ifmmode R_{\rm vir}^{\rm h} \else 
    $R_{\rm vir}^{\rm h}$\fi}}

\newcommand{\Ncl}{{\ifmmode N_{\rm cl} \else $N_{\rm cl}$\fi}}
\newcommand{\ncl}{{\ifmmode n_{\rm cl} \else $n_{\rm cl}$\fi}}
\newcommand{\phicl}{{\ifmmode \phi_{\rm cl} \else $\phi_{\rm cl}$\fi}}
\newcommand{\phicltot}{{\ifmmode \phi_{\rm cl}^{\rm tot} \else 
    $\phi_{\rm cl}^{\rm tot}$\fi}}
\newcommand{\Beff}{{\ifmmode B_{\rm eff} \else $B_{\rm eff}$\fi}}

\newcommand{\aap}{Astron.~Astroph.}

\newcommand{\apss}{Astrophysics and Space Science}

\newcommand{\jgr}{J.~Geophys.~Res.}

\newcommand{\mnras}{MNRAS}

\begin{document}


\title{
Voyager 1 $e^\pm$ Further Constrain Primordial Black Holes as Dark Matter
}
\author{Mathieu Boudaud}
\email{boudaud@lpthe.jussieu.fr}
\author{Marco Cirelli}
\email{marco.cirelli@gmail.com}
\affiliation{Laboratoire de Physique Th\'eorique et Hautes Energies (LPTHE), CNRS \& Sorbonne Universit\'e, 4 Place Jussieu, Paris, France}

\begin{abstract}
Primordial black holes (PBHs) with a mass $M \lesssim {10^{17}}$g are expected to inject sub-GeV electrons and positrons in the Galaxy via Hawking radiation. These cosmic rays are shielded by the solar magnetic field for Earth-bound detectors, but not for Voyager 1, which is now beyond the heliopause. We use its data to constrain the fraction of PBHs to the dark matter in the Galaxy, finding that PBHs with $M<10^{16}$g cannot contribute more than 0.1\% (or less for a log-normal mass distribution). Our limits are based on local Galactic measurements and are thus complementary to those derived from cosmological observations.


\end{abstract}

\maketitle


{\textit{Introduction} ---}
One of the most pressing questions in current cosmology is the nature of the dark matter (DM) that constitutes about 26\% of the total energy-matter content of the Universe~\cite{Ade:2015xua}. 
A large part of the theoretical and experimental efforts in the past decades have focused on explaining it as a new particle beyond the standard model (SM) of particle physics,
%
which, however have not shown up yet. The initial alternative proposal~\cite{ZeldovichPBH, Hawking:1971ei} that DM could instead consist of primordial black holes (PBHs) has therefore recently and deservedly come back to the attention of the community
(see~\cite{Carr:2009jm, Khlopov:2008qy,Carr:2016drx} for milestone reviews). 

These objects would be generated in the early Universe when sufficiently large density perturbations in the primordial plasma collapse gravitationally. If they are formed early enough, the material of which they are made is subtracted very early on from the baryonic budget, and therefore they are not subject to the cosmological constraints from primordial nucleosynthesis and the cosmic microwave background (CMB).
A number of possible mechanisms exist which could generate the needed large primordial fluctuations, invoking more or less exotic cosmological inflationary ingredients~\cite{GarciaBellido:1996qt, Jedamzik:1999am, Suyama:2006sr, Kohri:2012yw, Kawasaki:2012wr, Bugaev:2013vba, Drees:2011yz, Drees:2011hb, Kawaguchi:2007fz, Kohri:2007qn, Clesse:2015wea,Ballesteros:2017fsr} or just relying on SM ones~\cite{Ezquiaga:2017fvi,Espinosa:2017sgp}, albeit in peculiar configurations~\cite{Gross:2018ivp,Espinosa:2018euj}. 
In general terms, the expected mass of a PBH is connected to the time $t$ at which it was created, $M \sim c^3t/G \simeq 10^{15} (t/10^{-23} {\rm s}) {\rm g} \simeq 5 \times 10^{-19} (t/10^{-23}{\rm s}) M_\odot$, where $c$ is the speed of light, $G$ is the Newton constant, and $M_\odot \simeq 2 \times 10^{33}$ g is the mass of the Sun.
%
%
This relation illustrates that a very large range of masses is possible. PBHs created at the Planck time ($10^{-43}$ s) would have a Planck mass ($10^{-5}$g), while those generated just before big-bang nucleosynthesis ($t \sim 1$ s) could have a mass of $\sim 10^5 M_\odot$, comparable in size to the supermassive BHs at the center of current galaxies. Moreover, realistic production mechanisms predict not just a unique mass for all PBHs but rather an extended mass function.

This very large mass range is subject to a number of constraints. Broadly speaking, large masses ($10^3 M_\odot$ and up) are bound by dynamical constraints~\cite{LaceyOstriker,DeRujula:1991vy,Carr:1997cn,Brandt:2016aco,Koushiappas:2017chw,Yoo:2003fr,Quinn:2009zg,Monroy-Rodriguez:2014ula}, such as the need of avoiding the disruption of observed binary stellar systems, globular clusters or the destabilization of the galactic disk or bulge. 
Large mass PBHs also accrete significant amounts of material, emitting radiation (x rays and radio) that is constrained by current observations~~\cite{Gaggero:2016dpq,Inoue:2017csr} and by the CMB~\cite{Ricotti:2007au,Ali-Haimoud:2016mbv,Poulin:2017bwe}. 
A wide range of intermediate masses ($\sim 10^{17} {\rm g} \to 10^{35} {\rm g}$) are constrained by strong lensing measurements~\cite{Alcock:2000ph,Wyrzykowski:2010mh,Wyrzykowski:2011tr,Tisserand:2006zx,Allsman:2000kg,Mediavilla:2009um,Griest:2013esa,Griest:2013aaa,Novati:2013fxa,Niikura:2017zjd} of different sources (stars, either in the Magellanic clouds or in Andromeda, or gamma-ray bursts), as well as by pulsar timing arrays using Shapiro time delay~\cite{Schutz:2016khr}.
The lower portion of this range might also be constrained by neutron star survival arguments~\cite{Capela:2013yf,Graham:2015apa}.
Whether some windows still exist in which PBHs (of fixed mass or distributed on a range of masses) can constitute 100\% of the DM is currently subject to an intense debate. 
Finally, very small masses ($ \lesssim 4 \times 10^{14}$ g) are ruled out because PBHs, like any BH, are believed to be subject to Hawking evaporation~\cite{Hawking:1974rv,Hawking:1974sw}, which would have made them disappear by now.


In this work, we are particularly interested in the mass range above the evaporation limit ($4 \times 10^{14}$ g) and below the lowest lensing limit ($10^{17}$ g). In this range, PBHs are Hawking evaporating right now, emitting particles with a characteristic spectrum centered around tens of MeV. Indeed, constraints have been derived in the past using extragalactic gamma-ray background (EGB) observations~\cite{Page:1976wx,Carr:1998fw,Barrau:2003nj,Carr:2009jm}.  While powerful, such constraints do not test the local DM density but rather its average extragalactic distribution. Moreover, they are subject to (mild) uncertainties related to the spectral index of extragalactic photons~\cite{Carr:2017jsz}.
Limits derived from the Galactic gamma-ray background (GGB) are also relevant for masses smaller than $\sim 10^{15}$g~\cite{Lehoucq:2009ge,Carr:2016hva}.
In the same range of masses, recent bound have been derived using Planck data~\cite{Stocker:2018avm} as well as the latest {EDGES} measurements of the 21 cm absorption at high redshift~\cite{Clark:2018ghm}. The former are subdominant with respect to the EGB ones, while the latter could be stronger. Since, however, they are still subject to large uncertainties, we will mostly compare with the EGB and GGB.
 
Charged particles such as antiprotons, electrons and positrons have also been considered in the past~\cite{Carr:1976zz,Turner:1981ez,Kiraly:1981ci,Halzen:1991uw,Barrau:1999sk}. The main difficulty with them is that, at the relevant sub-GeV energies, charged cosmic rays are strongly affected by the sphere of influence of the Sun, which significantly complicates the picture. The access to low energy is instead particularly important since, as per Hawking radiation, the larger the PBH mass, the less energetic the emitted particles. 

The crucial observation in this work is that this limitation is now overcome by the fact that the Voyager 1 spacecraft has recently crossed the heliopause threshold, thereby becoming capable of collecting low-energy electrons and positrons~\cite{Voyager-1, 2016ApJ...831...18C}, possibly emitted by the evaporating PBHs. This will allow us to impose novel constraints, which, in contrast to the gamma-ray ones, are based on local measurements.
In addition, we will consider the  data collected by {AMS-02}~\cite{Aguilar:2014mma}. These cover a higher energy range, starting at about 0.5 GeV. They will therefore be relevant for models where the $e^\pm$ are significantly accelerated during the propagation process.


The rest of this Letter is organized as follows. We first review the production of $e^\pm$ from PBH evaporation and their propagation in the local Galactic environment. Then, we derive the constraints comparing with the experimental data, both for a unique PBH mass and for a mass distribution. We also compare with existing constraints. We finally briefly discuss the results and conclude. 

\medskip
{\textit{Methodology} ---}
A BH with mass $M$ has a temperature~\cite{Hawking:1974rv,Hawking:1974sw}
\beq
T = \frac{1}{8 \pi  G M} \simeq 1.06 \, \left( \frac{10^{13} \, \rm{g}}{M} \right) \, \rm{GeV},
\eeq
with 
$\hslash = c = k_{\rm B} = 1$. BHs lose mass radiating particles at a rate
%
\beq
\label{eq:dM_dt_BH}
\frac{dM}{dt} = -5.34 \times 10^{25} f(M) \left( \frac{ \rm g}{M} \right)^2 \rm g / s,
\eeq
where $f(M)$ is the number of emitted particle species normalized to unity for $M \gg 10^{17}$ g. The spectrum of emitted $e^\pm$ is 
%
\beq
\frac{dN_{e}}{dt dE} = \frac{\Gamma_{e}}{2 \pi} \left[\exp \left(\frac{E}{T}\right) +1 \right]^{-1},
\eeq
where $\Gamma_e$ is the electron absorption probability, which, in the geometric optics limit (high energy), reads $\Gamma_e \simeq 27 G^2 M^2 E^2$~\cite{MacGibbon:1990zk}. Assuming that PBHs constitute all of the DM, the number of electrons injected at the position $\vecx$ in the Galaxy per unit of time, energy and volume is
\beq
Q(E, \vecx) = \frac{\rho(\vecx)}{\rho_\odot} \int \limits_{M_{\rm inf}}^{+\infty} \!\!\!\! dM  \; \frac{g(M)}{M} \frac{dN_e}{dt  dE},
\eeq
where $\rho(\vecx)$ is the DM density and $g(M) = M dN_{\rm PBH} / dM dV$ is the mass distribution of PBHs normalized to $\rho_\odot$. 
We consider two spherically symmetric DM halos: a Navarro-Frenk-White (NFW)~\cite{NFW:1997}  halo scaling like $1/r$ in the center of the Galaxy and a cored halo  featuring a constant density at small galactocentric radii. 
We use the kinematically constrained parameters provided in Table 6 of Ref.~\cite{2017MNRAS.465...76M}, where the DM density at the Sun position $R_\odot \simeq 8.2$ kpc  is $\rho_\odot \simeq 0.4$ GeV/cm$^3$, and, for the cored profile, the core radius is 7.7 kpc.

\smallskip

The transport of cosmic rays (CRs) $e^{\pm}$ in the Galaxy is described by a phenomenological diffusion model~\cite{1964ocr..book.....G, 1974Ap&SS..29..305B, Moskalenko:1997gh, Delahaye:2008ua, 2010A&A...524A..51D, Boudaud:2014dta} with a rigidity ($\mathcal{R}$) -dependent diffusion coefficient $K(\mathcal{R}) = K_0 \, \beta \, (\mathcal{R}/\rm{GV})^\delta$.
CR $e^{\pm}$ lose energy through synchrotron emission and inverse Compton scattering on the interstellar radiation field as well as interacting with the gas of the interstellar medium (ISM) (ionization, Coulomb interaction and bremsstrahlung).
They also undergo convection because of the galactic wind (leading to additional energy losses because of adiabatic expansion) and diffusive reacceleration induced by the Alfv\'en waves propagating in the interstellar plasma. The velocity of the galactic wind is assumed to be constant in modulus, $\vec{V}_c={\rm sgn}(z)\,V_c \vec{e}_z$, and the reacceleration is linked to the spatial diffusion through $D(\mathcal{R}) \propto V_a^2/K_0(\mathcal{R})$, the exact functional form depending on the propagation model adopted. 

The galactic geometry is described by the two-zone diffusion model (disk and diffusive halo) with the galactic radius $R = 20$ kpc and the vertical extension of the Galactic disk $2h = 0.2$ kpc.
The half-height $L$ of the diffusive halo is of the order of a few kiloparsecs and is discussed further below.
The transport equation is solved following the semianalytical method introduced in Ref.~\cite{Maurin:2001sj} and extended for sub-10~GeV $e^{\pm}$ while accounting for all propagation processes by the pinching method~\cite{Boudaud:2016jvj}.  

The propagation parameters are determined from the data of secondary to primary CR ratio. We make use of two benchmark sets of parameters, dubbed hereafter models {\it A} and {\it B}.
Model {\it A} is the model MAX of Refs.~\cite{Maurin:2001sj, Donato:2003xg}, where $L = 15$ kpc, $K_0=0.0765$ kpc$^2$/Myr, $\delta=0.46$, $V_a=117.6$ km/s, and $V_c=5$ km/s, derived from HEAO-3 (High Energy Astrophysical Observatory) boron/carbon (B/C) data~\cite{1990A&A...233...96E}. We checked in Ref.~\cite{Boudaud:2016jvj} that these parameters are consistent with {AMS-02} $e^+$ data.
%
%
For model {\it B}, we adopt the best fit parameters of Ref.~\cite{Reinert:2017aga}, which makes use of the new {AMS-02} B/C data~\cite{PhysRevLett.115.211101} to update the propagation parameters: We use $L= 15$ kpc, $K_0=0.125$ kpc$^2$/Myr, $\delta=0.507$, $V_a=0$ km/s, $V_c=1.3$ km/s, $\mathcal{R}_b = 275$ GV, $\Delta \delta = 0.157$, and $s = 0.074$, where $\mathcal{R}_b$,  $\Delta \delta$, and $s$ parametrize a break in the diffusion coefficient~\cite{Genolini:2017dfb}.
%
Reference~\cite{Reinert:2017aga} determined only the ratio $K_0/L$ but also obtained indications that $L > 4.1$ kpc from the {AMS-02} $e^+$ flux. Hence, we will vary $L = 4.1 \to 20$ kpc for model {\it B}. 
Model {\it A} features a strong diffusive reacceleration, possibly required by the antiproton flux measured by {AMS-02}~\cite{Reinert:2017aga}, while there is none in model {\it B}. Since sub-GeV CRs $e^{\pm}$ are more sensitive to reacceleration than CR nuclei, we anticipate that the flux of $e^{\pm}$ produced by radiating PBHs will be drastically different for {\it A} and {\it B}.
The two models are therefore quite diverse and allow to quantify the impact of the CR propagation uncertainty on our results.  
%

%

\begin{figure}[!t]
\centering
\includegraphics[width = 0.495\textwidth]{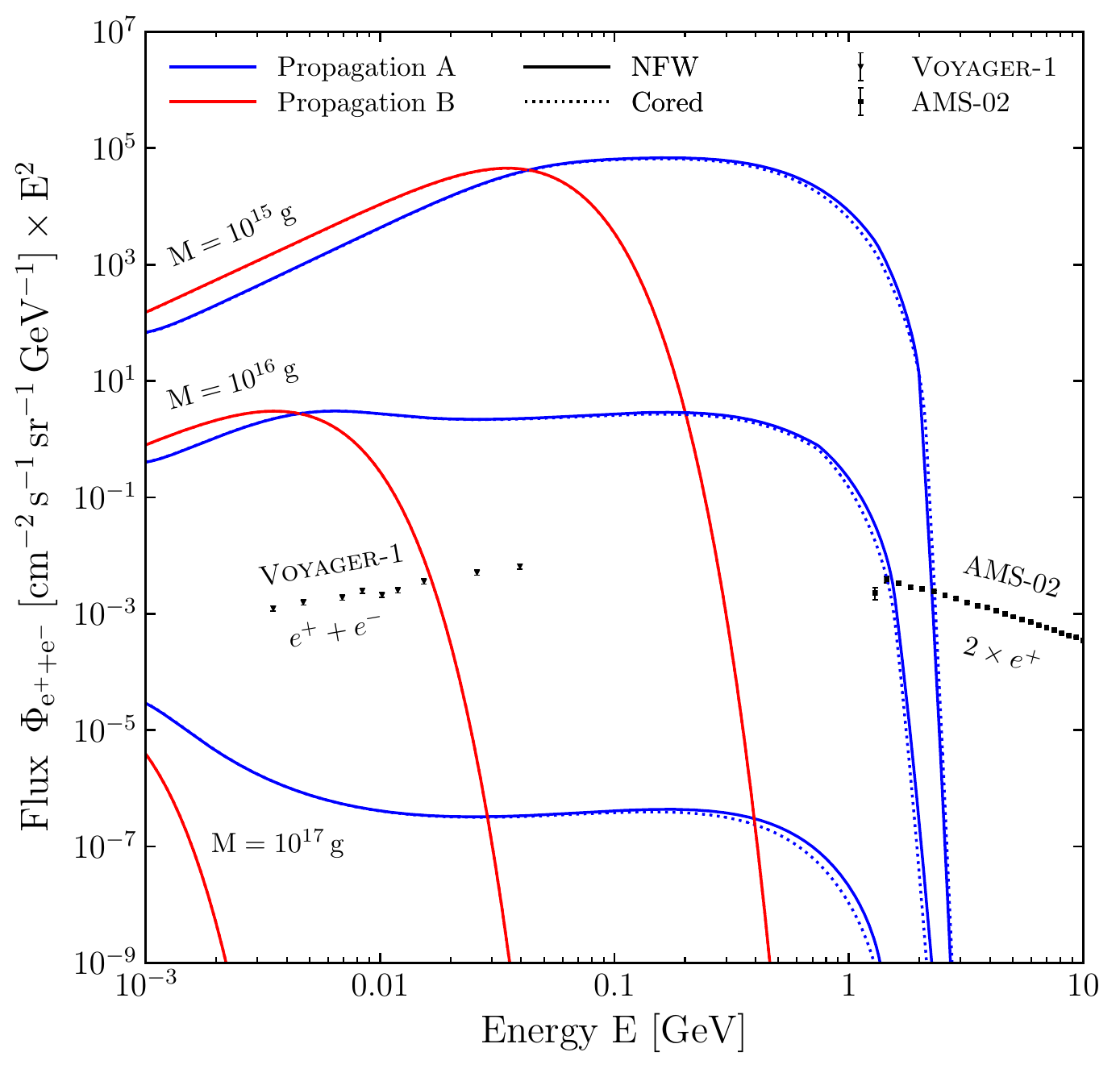}
\caption{\small Spectra of $e^\pm$ from PBH evaporation, after propagation in the Galaxy, for different PBH masses and with the indicated assumptions. The Voyager 1 and {AMS-02} data are also reported (the error bars are so small on this scale that they are included in the size of the point).}
\label{fig:spectra}
\end{figure}

\medskip
{\it Results ---}
Assuming all DM of the Galaxy is made of single-mass PBHs (monochromatic mass function), we represent in Fig.~\ref{fig:spectra} the flux of  $(e^{+} + e^{-})$ at the solar position produced by radiating PBHs with masses $10^{15}$, $10^{16}$ and $10^{17}$g.
%
%

The spectra obtained with model {\it B} drop very quickly above the PBH temperature. Indeed, since there is no diffusive reacceleration for {\it B}, the transport of sub-GeV $e^{\pm}$ is dominated by energy losses (mainly ionization of the ISM) and CR $e^{\pm}$ continuously cool down as they propagate. As a consequence, the bulk of $e^{\pm}$ measured at Earth are produced locally in a few kiloparsec radius sphere around the Sun, and their flux is approximatively given by
\beq
\Phi_{e^{\pm}}(E, \odot) \simeq \frac{c}{4 \pi  b(E)} \int \limits_E^{\infty} \!\! dE_{\rm s} \, Q(E_{\rm s}, \odot),
\label{eq:LE_limit}
\eeq
where $b(E)$ is the energy loss rate. For energies much smaller than the PBH temperature, the flux can be approximated by the analytical expression $\Phi_{e^{\pm}}(E, \odot) \simeq [ 11G \rho_{\odot}  \, \zeta(3) T^2 ] / [ 4 \pi^2 b(E) ]$, where $\zeta$ is the Riemann function. 
Therefore, the $e^{\pm}$ spectrum follows the energy dependence of the energy loss rate and scales as $T^2$ (or, equivalently $M^{-2}$).
%
Figure~\ref{fig:spectra} shows that Voyager 1 data probe PBHs with masses $M \lesssim 10^{16} \, \rm g$. 

The situation is different for model {\it A} since a fraction of sub-GeV $e^{\pm}$ gains energy from the diffusive reacceleration and populates the spectrum above the PBH temperature. This is remarkable, as it means that CR detectors can be sensitive to $e^{\pm}$ with energies above the maximum energy at which they had been injected in the Galaxy, namely, the PBH temperature. In this specific situation, even {AMS-02} is sensitive to signals produced by PBHs with $M \lesssim 10^{16}$g. Note that the $e^\pm$ measured by {AMS-02} are affected by the solar magnetic activity. We can nevertheless reconstruct the interstellar flux using the force field approximation~\cite{1971JGR....76..221F}. We adopt a Fisk potential $\phi_{\rm F} = 830$ MV, the $3\sigma$ upper value from~\cite{2016A&A...591A..94G}, to assess the minimal sensitivity of {AMS-02}.
In the following, however, we will make use of only the Voyager 1 data since they turn out to be more restrictive than the {AMS-02} ones for the PBHs abundance.
%
We finally note that the spectra are rather insensitive to the choice between an NFW and a cored profile. This is expected, since sub-GeV $e^{\pm}$ are produced in the  local environment, where the profiles are similar. Constraints derived from $e^{\pm}$ on the PBH local abundance will therefore be very robust regarding the uncertainty on the DM halo profile.

%


\begin{figure*}[!t]
\centering
\includegraphics[width = 0.495\textwidth]{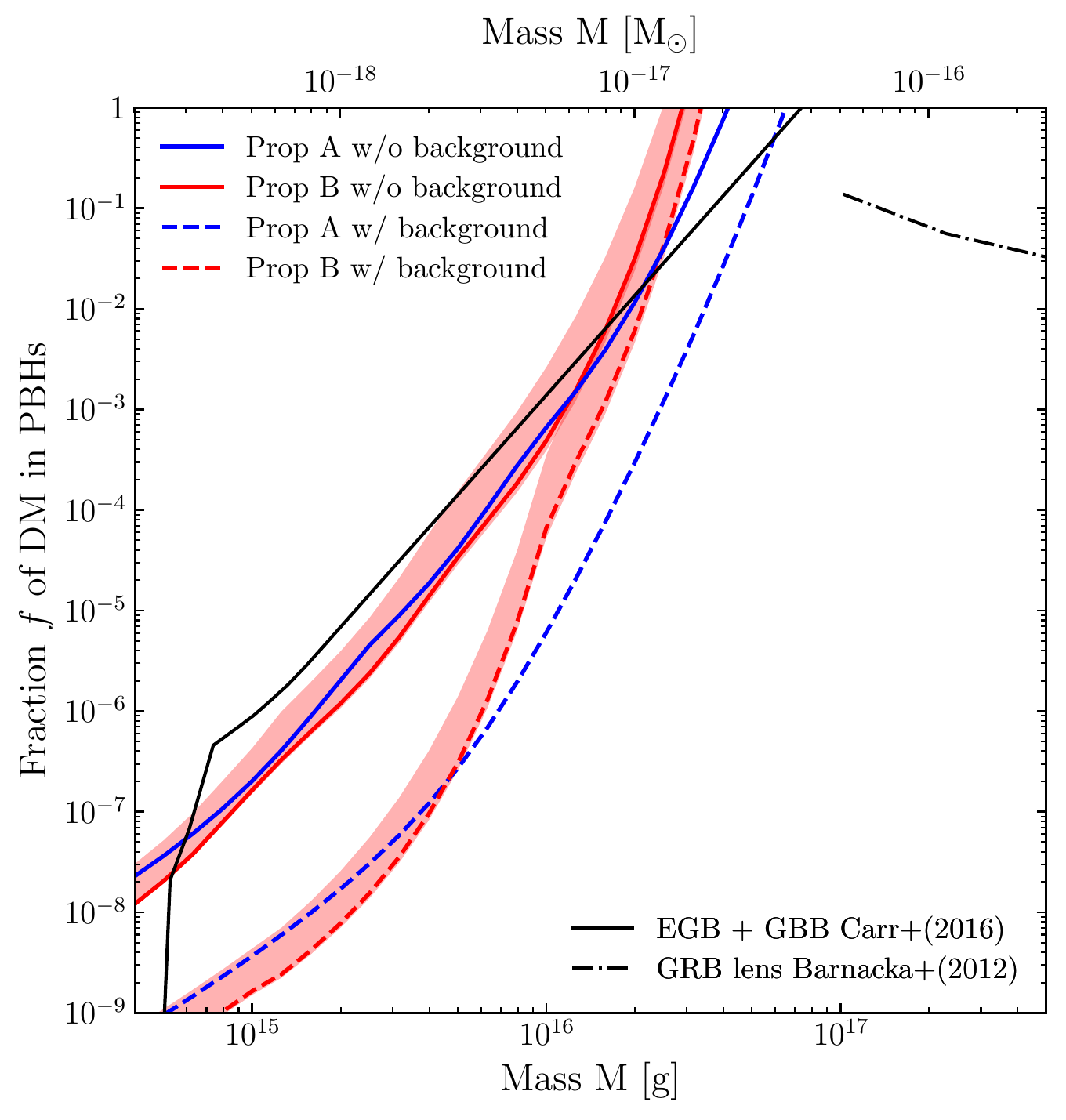}
\includegraphics[width = 0.495\textwidth]{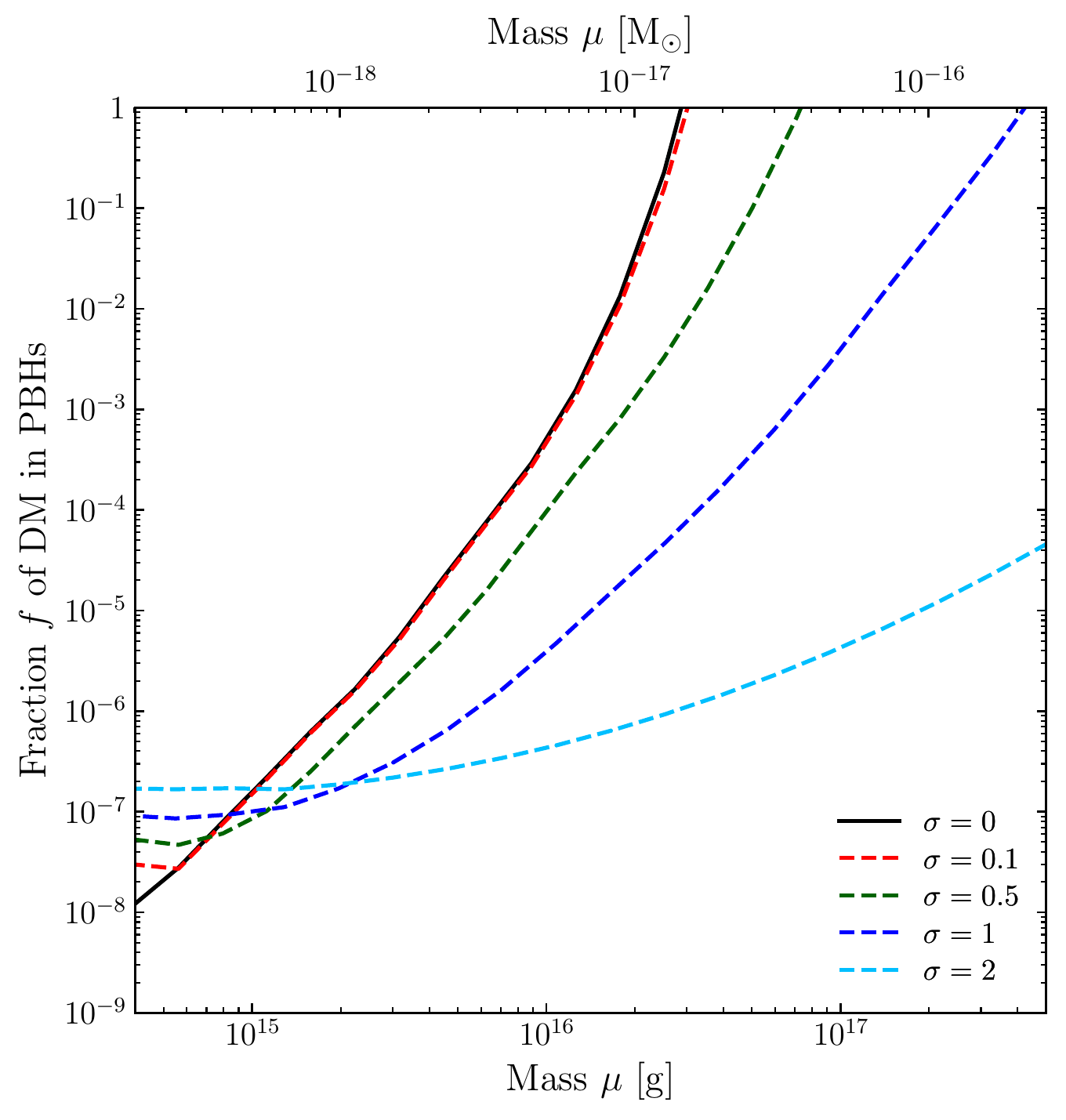}
\caption{\small Constraints on the fraction $f$ of PBHs to the DM as a function of the PBH mass, as obtained in this work (blue and red lines) and in related studies (black lines). The left panel assumes a single mass common to all PBHs, and the right panel assumes a log-normal mass distribution. The constraints in digital format are available upon request to the authors.}
\label{fig:constraints}
\end{figure*}

We thus use the Voyager 1 $e^{\pm}$ data to constrain the contribution of PBHs to the DM density in the Galaxy. 
The maximum fraction $f = \rho_{\rm PBH}/\rho_{\rm DM}$ is determined by requiring that the flux of $e^{\pm}$ emitted by PBHs does not overshoot any data point by more than $2\sigma$.
The limits for a monochromatic mass distribution are represented by the solid lines in the left panel in Fig.~\ref{fig:constraints}. 
The blue (red) solid line is obtained with model {\it A} (model {\it B}).
%
%
Regarding model {\it B}, ionization of the ISM dominates the transport of $e^\pm$ measured by Voyager 1 and thus the main uncertainty comes from the size $L$ of the diffusive halo (correlated with $K_0$ from the B/C analysis). Indeed, since sub-GeV $e^\pm$ almost do not lose energy in the diffusive halo, the larger the diffusive volume, the higher the signal from PBHs. This uncertainty affects the limits up to one order of magnitude as represented by the red band in Fig~\ref{fig:constraints}.
The uncertainty on the local dark matter density could also affect the limits at the level of 10\%~\cite{2017MNRAS.465...76M,Pato:2015dua}.
%
%
For both propagation models, PBHs with masses smaller than $10^{16}$g cannot contribute more than 0.1\% to the DM density of the Galaxy.


Up to now, we have not assumed any astrophysical background. There are, however, strong hints for the acceleration of Galactic $e^-$ by supernova remnants (SNRs)~(see, e.g.~\cite{Aharonian:2006dv}) and $e^{\pm}$ by pulsar wind nebulae (PWNe)~(see e.g:~\cite{Tibaldo:2017cyb}). Secondary $e^\pm$ may also contribute to the background, up to 10\% of the Voyager 1 flux~\cite{Boudaud:2016mos}.
Fitting the Voyager 1 data with a power law in energy, we find a spectral index of 1.31 (for $\chi^2_{\rm dof} = 10.1/9$).
Assuming negligible reacceleration, we found that this translates into a spectral index at injection of $\sim 2.1$ [this value can also be recovered directly from Eq.~(\ref{eq:LE_limit})], consistent with the value  predicted by diffusive shock acceleration simulations of SNRs and PWNe.
This thus suggests that these objects are likely responsible for the acceleration of the leptons measured by Voyager 1.
%
%
%
If we then assume a background for the Voyager 1 data modeled as a 1.31 power law, the room for a DM contribution significantly shrinks and the corresponding limits are represented by the dashed lines in Fig.~\ref{fig:constraints}.


Our limits 
without background are at the same level as the EGB ones for masses smaller than $10^{16}$g. 
On the other hand, taking into account a background probe, for $M \lesssim10^{16}$g, a fraction $f$ between 1 and 2 orders of magnitude smaller than the EGB. In the most constraining scenario (model {\it A} with a background) the limits almost reach the value $M \simeq 10^{17}$g probed by gamma-ray burst lensing~\cite{2012PhRvD..86d3001B}. (The recent analysis~\cite{Katz:2018zrn} investigates the dependance of these limits on effects related to the extended nature of the source as well as wave optics, making their robustness under debate.) 



So far, we have considered a single PBH mass.
Recent studies~\cite{Kannike:2017bxn, Carr:2017jsz, Bellomo:2017zsr, Calcino:2018mwh} suggest, however, that realistic production mechanisms result in an extended mass function. In some cases, the latter is well fit by a log-normal distribution
%
%
\beq
g(M) = \frac{\rho_{\odot}}{\sqrt{2\pi} \sigma M} \exp{\left(-\frac{\log^2(M/\mu)}{2 \sigma^2}\right)},
\eeq
where $\mu$ is the mass for which the density is maximal, $\sigma$ is the width, we have normalized to DM density at the Solar System position and
%
we cut at $4 \times 10^{14}$g, since all lighter PBHs have evaporated by today. 
The limits obtained in this case are represented in the right panel in Fig~\ref{fig:constraints} for different values of the width $\sigma$ in the range 0.1--2. We use here the propagation model {\it B} without any background for the Voyager 1 data. 
%
Considering this extended mass function enables us to further constraint the fraction $f$ with respect to a monochromatic distribution. This can be understood by the fact that the production rate of $e^{\pm}$ increases much more than the DM density as the PBH mass decreases. Therefore, the constraints are provided by the few light but very bright PBHs of the distribution.
For a width $\sigma$ larger than 1, Voyager 1 data exclude that PBHs can account for more than 1\% of the DM, for a central value of the log-normal distribution  $\mu \lesssim 10^{17}$g. 
Notice that EGB and GGB limits get also stronger with an extended mass function, leading to similar constraints~\cite{Carr:2017jsz,Carr:2016hva}.

We made use of Voyager 1 data to constrain the local abundance of PBHs, but there is, in principle, no reason preventing us from rather looking for a signal in the data. However, this requires a good understanding and modeling of the background of CRs $e^\pm$ in the sub-100 MeV energy range, which is beyond the scope of this Letter. Moreover, such a signal should be consistent with the EGB constraints reported in Fig.~\ref{fig:constraints}.

%


\medskip
{\it Conclusions ---}
In conclusion, we have made use of the capability of Voyager 1 of measuring the interstellar low-energy flux of CRs $e^{\pm}$ to constrain the contribution of PBHs to the DM in the Galaxy. 
We computed the flux of CRs $e^{\pm}$ Hawking radiated by PBHs using the fully general diffusion-convection-reacceleration model of propagation, with the most up-to-date parameters adjusted on the {AMS-02} data. Assuming that PBHs make up all the DM of the Galaxy, we found that Voyager 1 is sensitive to a signal from PBHs with $M \lesssim 10^{16}$g. {AMS-02} is also sensitive to PBHs with $M \lesssim 10^{16}$g for a propagation model with strong diffusion reacceleration. We therefore  constrained the fraction of PBHs to the DM density to be smaller than 0.1\% for $M \lesssim 10^{16}$g. We also showed that considering a log-normal mass distribution (as predicted by inflationary models) significantly improves the constraints. Our limits are competitive with those derived from cosmological observations and they are even better below $10^{16}$g when assuming an astrophysical background for the Voyager 1 data. 
These limits are robust regarding the DM distribution in the Galaxy and they are not affected by solar activity, precisely because Voyager 1 data have been collected beyond the heliopause. 
We estimate the propagation uncertainty on our limits to be around one order of magnitude. We emphasize that these new limits are based on local measurements and do not depend on any cosmological parameters. PBH clustering does not affect our results since the signal depends only on their density averaged on large scales in the Galaxy. They are therefore fully complementary to other limits derived from cosmological observations.


\section*{ACKNOWLEDGMENTS}
{We thank Bernard Carr, Alan Cummings, Julien Lavalle, Pierre Salati and Marco Taoso for useful discussions, and we thank Filippo Sala for an early collaboration on this work. 
M.B.~and M.C.~acknowledge the hospitality of the Institut d'Astrophysique de Paris (IAP) where a part of this work was done. 
This work is supported by the European Research Council ({\sc Erc}) under the EU Seventh Framework Program (FP7/2007-2013) / {ERC} Starting Grant (Agreement No 278234 -- {"NewDark"} project). 
It has been done in part within the Labex ILP (reference ANR-10-LABX-63) part of the Idex SUPER, and received financial state aid managed by the Agence Nationale de la Recherche, as part of the program {\it Investissements d'avenir} under the reference ANR-11- IDEX-0004-02.}

\bibliography{PBHs_Voyager}



\end{document}